\documentclass[nohyper]{JHEP3}

\usepackage{amsmath,amssymb}
\usepackage{graphicx}
\usepackage{cite}
\usepackage{amssymb}
\usepackage{epsfig}
\usepackage{subfigure}
\newcommand{\be}{\begin{equation}}
\newcommand{\ee}{\end{equation}}
\newcommand{\bea}{\begin{eqnarray}}
\newcommand{\eea}{\end{eqnarray}}

\providecommand{\mislash}[1]{#1 \mspace{-10.0mu} \slash}

\providecommand{\proarrow}[0]{\rightarrow}

\providecommand{\dif}[0]{\mathrm{d}}

\providecommand{\proname}[2]{#1 \proarrow #2}
\providecommand{\lrproname}[2]{#1 \leftrightarrow #2}

\providecommand{\miim}[1]{{\rm Im} \left[ #1 \right]}

\providecommand{\g}[2]{\gamma^{#1}_{#2}}

\providecommand{\gp}[2]{\gamma^{#1 \; '}_{#2}}

\title{Leptogenesis without violation of $B-L$}
\author{M.~C.~Gonzalez-Garcia\\
  Instituci\'o Catalana de Recerca i Estudis Avan\c{c}ats (ICREA), \\
  Departament d'Estructura i Constituents de la Mat\`eria and 
Institut de Ciencies del Cosmos,  
Universitat de Barcelona,\\
  Diagonal 647, E-08028 Barcelona, Spain\\
{\rm and:}  C.N. Yang Institute for Theoretical Physics\\
  State University of New York at Stony Brook\\
  Stony Brook, NY 11794-3840, USA,\\
  E-mail: \email{concha@insti.physics.sunysb.edu}}
\author{J. Racker\\
  Departament d'Estructura i Constituents de la Mat\`eria,
and Institut de Ciencies del Cosmos \\  
  Universitat de Barcelona,
  Diagonal 647, E-08028 Barcelona, Spain\\
  E-mail: \email{racker@ecm.ub.es}}
\author{N. Rius\\
Depto.\ de F\'{\i}sica Te\'orica and
IFIC, Universidad de
Valencia-CSIC \\ 
Edificio de Institutos de Paterna, Apt. 22085, 46071 Valencia,
Spain\\
E-mail: \email{nuria@ific.uv.es} }
\abstract{We study the possibility of generating the observed
baryon asymmetry via leptogenesis in the decay of heavy Standard Model 
singlet fermions which carry lepton number,  
in a framework without Majorana masses above the electroweak scale.  
Such scenario does not contain any source of total 
lepton number violation besides the Standard Model sphalerons,
and the baryon asymmetry is generated by the interplay of
lepton flavour effects and the sphaleron decoupling 
in the decay epoch.
}
\preprint{%
  IFIC/09-44 \\
  FTUV-09-0909 \\
YITP-SB-09-27}

\begin{document} 
\section{Introduction}

One of the most appealing and natural mechanisms for generating the 
tiny neutrino masses experimentally allowed is the (type I) see-saw
mechanism \cite{ss}. This simple extension of the Standard Model (SM)
with two or three right-handed Majorana neutrinos provides also 
a very attractive origin for the observed Baryon Asymmetry of the 
Universe, via leptogenesis \cite{fy,review}.
A lepton asymmetry is dynamically generated in 
the out of equilibrium decay of the heavy Majorana neutrinos, and 
then partially converted into a baryon asymmetry due to  
($B+L$)-violating non-perturbative sphaleron interactions \cite{krs}.

Unfortunately, a direct test of the see-saw mechanism is in general 
not possible. If the Dirac neutrino masses are similar to the other 
SM fermion masses, as naturally expected, the Majorana neutrino masses 
turn out to be of order $10^8-10^{16}$ GeV, so the 
right-handed (RH) neutrinos can 
not be produced in the LHC or future colliders, neither lead to other 
observable effects such as  lepton flavour violating processes. 
On the other hand,  
although the see-saw mechanism can also work with Majorana masses as low as 
100 GeV, which in principle are within the energy reach of the LHC, 
the smallness of the light neutrino masses generically implies in
this case tiny neutrino Yukawa  couplings and as consequence negligible 
mixing with the active neutrinos, leading also to unobservable 
effects.

In order to obtain a large active-sterile neutrino mixing, some 
cancellations must occur in the light neutrino mass matrix. 
In this context, the small neutrino masses are not explained 
by a see-saw mechanism, but rather by an ``inverse see-saw'' \cite{iss},
in which the global lepton number symmetry $U(1)_L$ is slightly 
broken by a small parameter, $\mu$. In other words, the 
fine-tuned cancellations required in the light neutrino 
masses are not accidental, but due to an approximate symmetry. 
The smallness of the parameter $\mu$ is
protected from radiative corrections (even without supersymmetry),
since in the limit $\mu \rightarrow 0$ a larger symmetry is 
realized \cite{tHooft}.

Much attention has been devoted recently to this possibility, 
both in the context of LHC phenomenology \cite{LHC} and 
in leptogenesis \cite{resonant,majoron,ab}. A consequence of the 
slightly broken $U(1)_L$ symmetry is the existence of two 
strongly degenerate RH neutrinos, which combine to form
a quasi-Dirac fermion. This is interesting for leptogenesis, 
because it leads to an enhancement of the CP asymmetry,
avoiding the strong bounds which apply to hierarchical 
RH neutrinos \cite{di}, 
and allowing for successful leptogenesis at
much lower temperatures, $T \sim {\cal O}(1 \;\rm{TeV})$
\footnote{See also \cite{marta,hambye} for alternative 
(extended) models of leptogenesis with RH Majorana 
neutrinos at the
TeV scale and without resonant enhancement.}.

We focus here on a different scenario: we assume that the small lepton 
number violating effects responsible of light neutrino masses 
are negligible during the leptogenesis epoch, so $B-L$ is 
effectively conserved. This can be the case if global lepton number is 
broken spontaneously at a scale well below the electroweak 
phase transition \cite{cv1} or, even if lepton number is 
broken at high scales, it leads to a CP asymmetry too small to 
account for the observed baryon asymmetry.  
The main difference with previous approaches is that in
this framework the
RH neutrinos combine exactly into Dirac fermions, 
and the total CP asymmetry vanishes.
As a consequence, in order to generate a baryon 
asymmetry we have to rely on 
i) flavour effects and ii) sphaleron departure from
thermal equilibrium during the leptogenesis epoch.

Leptogenesis without neutrino Majorana masses, so-called 
``Dirac leptogenesis'', has already been considered in the 
literature \cite{Diraclepto}, but in a completely
different set-up. 
In Dirac leptogenesis global lepton  
number remains exactly unbroken (except for the SM sphaleron
interactions), so the light neutrinos are Dirac fermions
made of the SU(2) doublet $\nu_L$ and the singlet 
RH neutrino $\nu_R$.
Realistic models contain not only the SM plus  
RH neutrinos, but also additional heavy 
particles to generate 
a non-zero lepton number for left-handed particles and 
and an equal and opposite lepton number for 
right-handed particles 
in their CP violating decay.

This paper is organized as follows. In Sec.~2 we describe our
leptogenesis framework, the CP asymmetries produced in the  
heavy Dirac neutrino decay and the basic requirements to 
generate the baryon asymmetry. In Sec.~3 we write the
network of Boltzmann equations relevant for leptogenesis 
without Majorana masses. In Sec.~4 we present our results, both
in the resonant and non-resonant regimes, and we conclude in Sec.~5.

\section{The Framework}
\label{sec:iss}
Our starting point is that above the electroweak phase transition 
the relevant particle contents for leptogenesis is that of the SM 
plus a number of SM singlet Dirac fermions $N_i$.
Without loss of generality we can work on a basis in which 
$N_i$ are mass eigenstates. In this basis 
the Lagrangian  above the electroweak phase 
transition can be written as: 
\begin{equation}
\mathcal{L} = \mathcal{L}_{\text{SM}} + i \overline{N}_i
  \mislash{\partial} N_i - M_i \overline{N}_i
  N_i 
- \lambda_{\alpha i}\,{\widetilde h}^\dag\, \overline{P_R N_i} \ell_\alpha 
- \lambda^*_{\alpha i } \overline{\ell}_\alpha P_R N_i {\widetilde h},
\label{eq:lag}
\end{equation}
where $\alpha,i$ are family indices ($\alpha=e, \mu, \tau$ and 
$i=1,2,3, \dots$), $\ell_\alpha$ are the leptonic $SU(2)$ doublets, 
$h=(h^+,h^0)^T$ is the Higgs field ($\widetilde h =i\tau_2 h^*$, with $\tau_2$
Pauli's second matrix) and $P_{R,L}$ are the chirality projectors. 

A quantitative illustration of the proposed scenario which can
also account for the  low energy neutrino phenomenology can
be found in the context of the inverse see-saw mechanism. 
In this type of models \cite{iss}, 
the lepton sector of the 
Standard Model is extended with 
two electroweak singlet two-component leptons per generation, i.e., 
\be
\ell_i = 
\left( \begin{array}{c}
\nu_{ L i} \\ 
e_{L i}
\end{array} \right),e_{R i}, \nu_{R i}, s_{L i} \; .
\ee

In the original formulation, the singlets $s_{L i}$ were 
superstring inspired E(6) singlets, in contrast to the right-handed
neutrinos $\nu_{R i}$, which are in the spinorial representation. 
More recently this mechanism has also arisen in the context of 
left-right symmetry \cite{alsv} and SO(10) unified models \cite{barr}.

At zero temperature the  (9 $\times$ 9) mass matrix of the neutral 
lepton sector in the $\nu_L, \nu_R^c, s_L$ basis is given by 
\be
\label{eq:M}
\cal{M} = 
\left( \begin{array}{ccc}
0    &    m_D    &    0 \\
m_D^T  &    0    &    M \\
0    &    M^T    &    \mu  \\
\end{array} \right) \; ,
\ee
where $m_D$ and $M$ are arbitrary 3 $\times$ 3 complex matrices in flavour 
space with 
\be
m_D\equiv \lambda_{\alpha i}  \, v  \; ,
\ee
and $v=174$ GeV being the Higgs vacuum expectation value. Moreover 
$\mu$ is a $3\times 3$ complex symmetric matrix. 

The matrix $\cal{M}$ can be diagonalized 
by a unitary transformation, leading to nine mass eigenstates: 
three of them correspond to the observed light neutrinos,
and the other six are heavy Majorana neutrinos. 

In this ``inverse see-saw'' scheme, assuming  $m_D,\mu \ll M$
the effective Majorana mass matrix for the light neutrinos is 
approximately given by 
\be
\label{mnu}
m_\nu = m_D {M^T}^{-1}\mu M^{-1}m_D^T \ ,
\ee
while the three pairs of two--component heavy neutrinos combine 
to form three quasi-Dirac fermions with  masses of order $M$. The 
admixture among singlet and doublet $SU(2)$ states 
(and the corresponding violation of unitarity in the light lepton sector) 
is of order $m_D/M$ and can be large \cite{bsvv,cv2,mfss}.  
This is so because  although $M$ is a large mass scale suppressing 
the light neutrino masses,
it can be much smaller than in the standard type-I see-saw scenario 
since light neutrino masses in Eq.~(\ref{mnu})
are further suppressed by the small ratio $\mu/M$.
Thus, in this scenario, for $M$ as low as the electroweak scale 
the only bounds on the Yukawa couplings are those arising from   
constraints on  violation of weak universality, lepton flavour violating 
processes and collider signatures \cite{unitlim}.

It is important to notice that in the $\mu \rightarrow 0$ limit a 
conserved total lepton number can be defined. This can be easily seen 
if, together with the standard lepton number $L_{SM}=1$ for the SM leptons 
we assign  a lepton number $L_N = 1$  to the singlets 
$\nu_{R i}$ and $s_{L i}$.  
With this assignment the  mass matrix \eqref{eq:M} with $\mu=0$ 
conserves  $L \equiv L_{SM}+L_N$.
Then, the three light neutrinos are 
massless Weyl particles
while
the six heavy neutral leptons combine exactly into three Dirac fermions, 
$N_i$, which above the electroweak scale are given by:
\begin{equation}
\label{Diracn}
N_i=s_{L i} + \nu_{R i} \;.
\end{equation}

The smallness of the $\mu$ term can be easily understood if 
the total lepton number is spontaneously broken by a vacuum 
expectation value $\langle \sigma \rangle$, with 
$\mu = f \langle \sigma \rangle$ \cite{cv1}. In this case, 
light neutrino masses are a consequence of 
total lepton number being broken at an energy scale much lower than 
the electroweak scale  $\langle \sigma\rangle \ll v$, 
and $\mu$ vanishes exactly at the heavy neutrino decay epoch.
This scenario introduces one extra scalar singlet which couples 
with  $s_L$ and the SM Higgs as
\be
\mathcal{L}_{int} = - \frac 1 2 f_{ij} \overline{s^c_{L_i}} \sigma s_{L_j}
+ \lambda |h|^2 |\sigma|^2  \;,  
\ee
and therefore can affect our results when
considered in the framework of the inverse see-saw.
In principle, there is a new Dirac neutrino decay channel, 
$N_i \rightarrow N_j \sigma$, which could be relevant for leptogenesis; 
in practice, 
as we will see the present mechanism only works for very degenerate 
heavy singlets, $M_1 \simeq M_2$, therefore this channel  is 
phase-space suppressed and our analysis will remain valid.

In this framework, if $\mu$ is effectively zero at the leptogenesis
epoch,  all processes conserve $B$ and $L$ at the perturbative level. 
On the other hand, the sphalerons violate $B+L_{SM}$ but conserve 
$B-L_{SM}$ and,
since the new heavy leptons are SM singlets, they 
do not change $L_N$. Therefore the SM sphaleron processes also conserve 
$B-L$. In brief $B-L$ is conserved by all the interactions 
of the model on scales above  $\langle\sigma\rangle$. 

Thus, one is effectively working in the limit in which 
the three light neutrinos are 
massless, the heavy ones combine into three Dirac fermions
given by Eq.~\eqref{Diracn} 
and the relevant interactions are those  in Eq.~\eqref{eq:lag}.

Indeed, if leptogenesis occurs via the decay of heavy Standard Model 
singlets in a framework without Majorana masses above the electroweak scale,
Eq.~\eqref{eq:lag} has the relevant information. Thus our results will hold 
whether light neutrinos acquire masses by the
inverse see-saw mechanism with the $\mu$ term generated at low 
scales as described above or by some other mechanism, as long as
it does not imply the presence of new states relevant for leptogenesis.

\subsection{The CP Asymmetries}
\label{sec:CP}

One important ingredient that determines the baryon asymmetry generated
in thermal leptogenesis in this scenario is the CP asymmetry 
produced in the decays of the heavy Dirac neutrinos $N_i$ into leptons 
of flavour $\alpha$, $\epsilon_{\alpha i}$:
\begin{equation}
\epsilon_{\alpha i} \equiv 
\frac{\displaystyle \Gamma(\proname{N_i}{\ell_\alpha h})
- \Gamma(\proname{\bar N_i}{\bar \ell_\alpha \bar h})}
{\displaystyle \sum _\alpha
    \Gamma(\proname{N_i}{\ell_\alpha h} )+ \Gamma(\proname{\bar N_i}{\bar
    \ell_\alpha \bar h})} \; .
\end{equation}
\FIGURE[t]{
\centering
\includegraphics[width=0.8\textwidth]{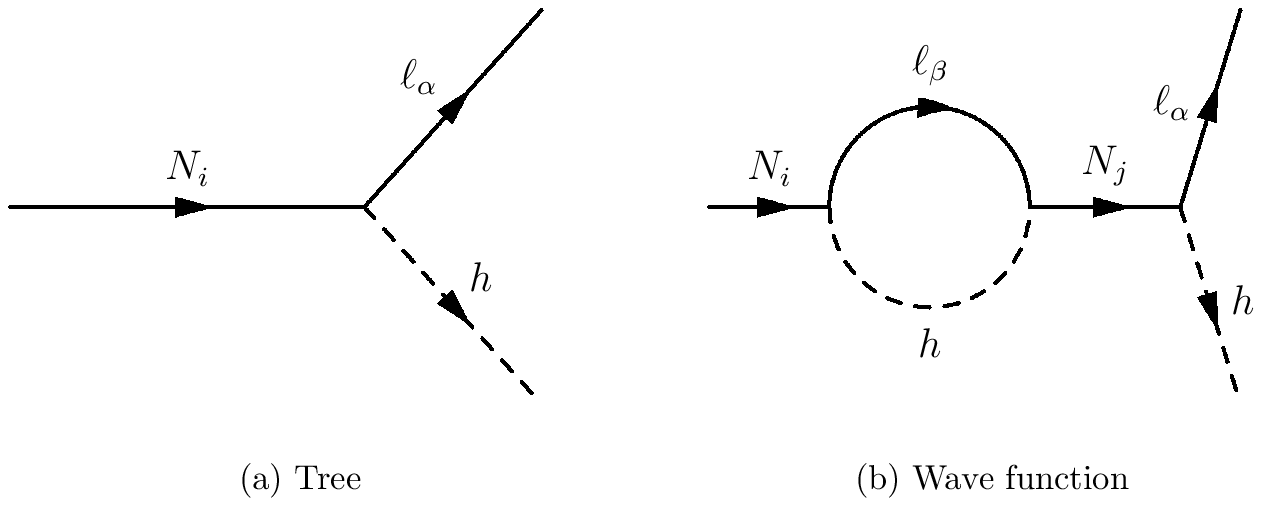}
\caption[]{The tree and one loop diagrams that contribute to the CP
asymmetry in decays when the heavy neutrinos are of Dirac type.
\label{fig:loop}}}

Because of the Dirac nature of $N_i$, 
the only one-loop contribution to the CP asymmetry arises 
from the interference of the tree-level and the self-energy 
one-loop  diagrams displayed in Fig.~\ref{fig:loop}, and it is given 
by \cite{crv,resonant} 
\footnote{We have already regularized the divergence at 
$M_i = M_j$ using the resummation procedure of \cite{resonant}.
In \cite{abp} a different calculation was performed, obtaining that the 
regulator of the singularity is $M_j \Gamma_j - M_i \Gamma_i$
instead of $M_i \Gamma_j$, but for the values of the widths 
that we consider ($\Gamma_j \gg \Gamma_i$), both results coincide.}
\begin{eqnarray}
\epsilon_ {\alpha i} &=& \frac{-1}{8 \pi (\lambda^\dag \lambda)_{ii}}  
\sum_{j \neq i} \frac{a_j-1}{(a_j-1)^2+g_j^2} 
\miim{\lambda_{\alpha j }^* \lambda_{\alpha i} 
(\lambda^\dag \lambda)_{i j}} \nonumber \\
&=&
 \frac{-1}{8 \pi}\sum_{j \neq i} \frac{a_j-1}{(a_j-1)^2+g_j^2}
(\lambda^\dag \lambda)_{jj}
\sqrt{K_{\alpha i}} \sqrt{K_{\alpha j}} \sum_{\beta \neq 
\alpha} \sqrt{K_{\beta i}}
\sqrt{K_{\beta j }} \, p_{\alpha\beta}^{ij} \; ,
\label{eq:epsi}
\end{eqnarray}
 where $a_j \equiv M_j^2/M_i^2$ ,
$g_j\equiv \Gamma_j/M_i$ and
\be
\Gamma_i=\frac{M_i}{8\pi} (\lambda^\dagger \lambda)_{ii}
\equiv \frac{1}{8\pi} \frac{\tilde m_i}{v^2}{M_i^2} 
\ee
is the decay width of $N_i$. 
In the last equation we have also introduced the effective mass $\tilde m_i$. 
Notice that the above CP asymmetry is only 
part of the usual wave function contribution for Majorana $N_i$, 
so that in the Dirac case   
the total CP asymmetry exactly vanishes
\begin{equation}
\epsilon_i \equiv \sum_\alpha \epsilon_{\alpha i}=0 \:, 
\end{equation} 
by CPT invariance and unitarity.
Thus in order for leptogenesis to occur we must be in a temperature
regime in which flavour effects~\cite{flavour, flavour2} are important.

In writing Eq.~\eqref{eq:epsi} we have expressed 
the Yukawa couplings in terms of the
projection coefficients $K_{\alpha i}$ 
(related to the absolute values of the Yukawas)
and some phases $\phi_{\alpha i}$ as 
\begin{equation}
\lambda_{\alpha i} = \sqrt{K_{\alpha i}} \sqrt{(\lambda^\dag \lambda)
_{ii}} e^{i \phi_{\alpha i}} \: ,
\end{equation}   
where
\begin{eqnarray}
K_{\alpha i } &=& \frac{\lambda _{\alpha i} 
\lambda_{\alpha i}^*}{(\lambda^\dag \lambda)_{ii}} \; ,  \\
p_{\alpha\beta}^{ij}&=&-p_{\beta\alpha}^{ij}=
-p_{\alpha\beta}^{ji}=\sin(\phi_{\alpha i}- \phi_{\alpha j}+
\phi_{\beta j}- \phi_{\beta i}) \; .
\end{eqnarray}
The factor $\frac{a_j-1}{(a_j-1)^2+g_j^2}$  can be resonantly enhanced to 
$M_j/2 \Gamma_j$ if $M_j-M_i\sim \Gamma_j/2$ \cite{resonant}. 
In this regime the resonant contribution to the asymmetry becomes
\begin{equation}
\epsilon^{res}_ {\alpha i}=  - \frac{1}{2}
\sqrt{K_{\alpha i}} \sqrt{K_{\alpha j}} \sum_{\beta \neq 
\alpha} \sqrt{K_{\beta i}}
\sqrt{K_{\beta j }} \, p_{\alpha\beta}^{ij} \; . 
\end{equation}

\subsection{The Generation of $B$: Basic Requirements}
\label{sec:requirements}
In the scenario that we are presenting $B - L_{SM} - L_N$ is
conserved by all processes. Since we want to explore the possibility
of generating the cosmological baryon asymmetry exclusively during the
production and decay of the heavy Dirac neutrinos, we assume that at
the beginning of the leptogenesis era $B-L_{SM} =0$ and $L_N=0$ (same
abundance of $N_i$ and $\overline N_i$)\footnote{In this work we will
always consider the case in which the heavy neutrinos are produced
thermally exclusively through their Yukawa interactions, so that their
abundance is null at the beginning of Leptogenesis. However we have
also studied cases in which their initial abundance is that of a
relativistic particle in equilibrium with the thermal bath (still
satisfying $L_N (\text{initial})=0$) and found no significant
differences.}. Therefore it is clear that
$B-L_{SM} = 0$ after the heavy neutrinos have disappeared. If the
sphalerons were still active after the decay epoch, the final
baryon asymmetry, being proportional to $B-L_{SM}$, would be zero.

Then, in order for leptogenesis to occur, one must be in the regime in
which the sphalerons depart from equilibrium (which occurs at $T \sim
T_f$) {\sl during the decay epoch}. In this case, as described in next
section, the baryon asymmetry freezes at a value $Y_B \propto
Y_{B-L_{SM}} (T = T_f)$, which in general is not null.

To go on we assume that the observed baryon asymmetry is generated
during the decay epoch of the lightest heavy neutrino, $N_1$.  For the
sake of concreteness we will assume that $M_3\gg M_2\geq M_1$ so the
maximum contribution to the CP asymmetry in $N_1$ decays is due to
$N_2$. Note that, in this scenario, for hierarchical masses the CP 
asymmetry is suppressed as $(M_1/M_2)^2$, instead of 
$M_1/M_2$ in the standard case with heavy Majorana  neutrinos. 
This contributes to the 
impossibility of generating enough $B$  
with hierarchical heavy neutrinos.
Moreover,  if $N_1$ and $N_2$ have similar masses, they will coexist
during the leptogenesis era, and there will be 
lepton flavour violating (but total lepton number conserving) washouts 
processes involving real $N_2$ as well as real $N_1$.
In what follows we will refer to the washout processes involving real or virtual heavy neutrinos as
lepton flavour violating washouts  (LFVW).

This implies that, compared to high scale leptogenesis
models with $M_i \gtrsim 10^{9}$~GeV, this electroweak scenario
typically suffers from very strong LFVW 
\footnote{ Similarly strong $\Delta L \neq 0$ washouts are also expected in 
standard (non-resonant)
leptogenesis with Majorana neutrino masses at the weak scale
\cite{marta,hambye}.}.
The argument goes as  follows: 
the intensity of the LFVW
associated with processes involving real or virtual $N_2$ is determined by
the adimensional parameter $\tilde m_2/m_*$
\footnote{The
quantity $m_*$ is the {\it equilibrium mass}, which is defined by the
condition $\tfrac{\Gamma_{N_i}}{H(T=M_i)}=\frac{\tilde m_i}{m_*}$,
where $H$ is the Hubble expansion rate, so that $m_*=\tfrac{16}{3
\sqrt{5}} \pi^{5/2} \sqrt{g_{*SM}} \frac{v^2}{m_{pl}} \simeq 1,08
\times 10^{-3}$~eV ($g_{*SM}$ is the number of Standard Model
relativistic degrees of freedom at temperature $T$ and $m_{pl}$ is the
Planck mass).} . 
For fixed values of the Yukawas of $N_2$ -- basically
for a fixed value of the CP asymmetry in $N_1$
decays--  $\tilde m_2/m_*$ scales as
$m_{pl}/M_2$. Therefore the LFVW will be larger the lower $M_2$.

Consequently in order to avoid a
complete erasure of the asymmetry generated in the processes involving
$N_1$, either the CP asymmetry has to be resonantly enhanced so that
the Yukawas of $N_2$ can be small while still having enough CP
asymmetry, or some of the projectors $K_{\alpha 2}$ have to be very
small to reduce the LFVW in some of the lepton flavours.

Altogether we find the following {\sl minimum} requirements:
\begin{itemize}
\item The mass of $N_1$ must be not far from the
sphaleron freeze-out temperature $M_i\lesssim {\cal O}(TeV)$. 
\item Either the CP asymmetry is resonantly enhanced or 
the Yukawa couplings of $N_2$ have a strong flavour hierarchy 
(i.e. $K_{\alpha 2} \ll 1$ for $\alpha = e, \mu$ or $\tau$). 
\item Even in the non-resonant case there cannot be a large hierarchy 
between $M_1$ and $M_2$ in order to
have as little suppression as possible
from the  $\frac{1}{a_2-1}$ factor. 
\end{itemize}

With these requirements we conclude that to quantitatively
determine the viability of this scenario we need to consider 
the evolution of the abundances of both $N_1$ and $N_2$ (as well
as the corresponding $\overline N_i's$) and the lepton flavour 
asymmetries. In order to do so we solve the set of Boltzmann equations
which we describe next.

\section{The Boltzmann Equations}
\label{be}
In writing the relevant Boltzmann Equations we first notice that, 
contrary to the case in which the heavy neutrinos are of Majorana
nature, the densities of $N_i$ and $\bar N_i$ can be different and 
enter separately in the Boltzmann equations. Thus, 
in general there is  an asymmetry between these two degrees of
freedom  which induces additional washout  of the lepton asymmetry. 
Moreover, the usual $\Delta L_{SM} =2$ processes 
mediated by the heavy neutrinos (like $\ell_\alpha h \rightarrow 
\bar{\ell}_\beta \bar{h}$, etc.) are absent, since total lepton 
number is perturbatively conserved. 
Therefore, the washout of the lepton asymmetries is due to 
$\Delta L_{SM}=0$ lepton flavour violating scatterings 
 mediated by the $N_i$
($\ell_\alpha h \rightarrow \ell_\beta h$, etc.), 
and 
$\Delta L_{SM} = -\Delta L_N = \pm 1$ reactions with one external $N_i$.

Considering all the $1 \leftrightarrow 2$ and $2 \leftrightarrow
2$ processes resulting from the Yukawa interactions of the heavy
neutrinos and the Yukawa interaction of the top quark,  
the evolution of the different densities is given by:
\begin{eqnarray}
- zHs \frac{\dif Y_{N_i+\bar N_i}}{\dif z} &=& \sum_\alpha
\{\lrproname{N_i}{\ell_\alpha h}\} + \{\lrproname{N_i \bar
  \ell_\alpha}{\bar Q_3 t}\} + \{\lrproname{N_i \bar t}{\bar Q_3
  \ell_\alpha}\} + \{\lrproname{N_i Q_3}{t \ell_\alpha}\}  \nonumber \\ 
&&  +
\sum_{\alpha, \beta, j \neq i} \{\lrproname{N_i \bar N_j}{\ell_\alpha
  \bar \ell_\beta}\} + \{\lrproname{N_i \ell_\beta}{N_j \ell_\alpha}
\}' + \{\lrproname{N_i \bar \ell_\alpha}{N_j \bar \ell_\beta} \}  
\label{eq:be1-1} \\ 
&& + \sum_{j \neq i} \{\lrproname{N_i \bar N_j}{h \bar h}\} +
\{\lrproname{N_i h}{N_j h }\}' + \{\lrproname{N_i \bar h}{N_j \bar
  h}\} \; , \nonumber \\
- zHs \frac{\dif Y_{N_i-\bar N_i}}{\dif z} &=& \sum_\alpha
(\lrproname{N_i}{\ell_\alpha h})+ (\lrproname{N_i \bar
  \ell_\alpha}{\bar Q_3 t}) + (\lrproname{N_i \bar t}{\bar Q_3
  \ell_\alpha}) + (\lrproname{N_i Q_3}{t \ell_\alpha}) \nonumber\\ 
& &+
\sum_{\alpha, \beta, j \neq i} (\lrproname{N_i \bar N_j}{\ell_\alpha
  \bar \ell_\beta}) + (\lrproname{N_i \ell_\beta}{N_j \ell_\alpha })'
+ (\lrproname{N_i \bar \ell_\alpha}{N_j \bar \ell_\beta}) \label{eq:be1-2}\\ 
& & +
\sum_{j \neq i} (\lrproname{N_i \bar N_j}{h \bar h}) + (\lrproname{N_i
  h}{N_j h })' + (\lrproname{N_i \bar h}{N_j \bar h}) \; , \nonumber\\ 
- zHs
\frac{\dif Y_{\Delta_\alpha}}{\dif z} &=& \sum_i
(\lrproname{N_i}{\ell_\alpha h}) + (\lrproname{N_i \bar
  \ell_\alpha}{\bar Q_3 t}) + (\lrproname{N_i \bar t}{\bar Q_3
  \ell_\alpha}) + (\lrproname{N_i Q_3}{t \ell_\alpha}) \nonumber\\ 
& & +
\sum_{i, j, \beta \neq \alpha} (\lrproname{N_i \bar N_j}{\ell_\alpha
  \bar \ell_\beta}) + (\lrproname{N_i \ell_\beta}{N_j \ell_\alpha })'
+ (\lrproname{N_i \bar \ell_\alpha}{N_j \bar \ell_\beta})  
\label{eq:be1-3}\\ 
& &+
\sum_{\beta \neq \alpha} (\lrproname{ h \bar h}{\ell_\alpha \bar
  \ell_\beta}) + (\lrproname{\ell_\beta \bar h}{\ell_\alpha \bar h}) +
(\lrproname{\ell_\beta h}{\ell_\alpha h})' \; ,
\nonumber
\end{eqnarray}
where $Y_X \equiv n_X / s$ is the number density 
of a single degree of freedom of the particle specie 
$X$ normalized to the entropy density
and $y_X \equiv (Y_X - Y_{\bar X})/Y_X^{eq}$ 
(to be used below) is the asymmetry density
normalized to the equilibrium density. With $Q_3$ and $t$ we denote 
respectively the third generation quark doublet and the top $SU(2)$ singlet. 
We have also defined $Y_{\Delta_\alpha} \equiv Y_B/3 - Y_{L_\alpha}$, 
where $Y_B$ is the baryon asymmetry and 
$Y_{L_\alpha}=(2y_{\ell_\alpha}+y_{e_{R\alpha}})Y^{eq}$
is the total lepton asymmetry in the flavour $\alpha$ 
(with $Y^{eq} \equiv Y_{\ell_\alpha}^{eq}
= Y_{e_{R\alpha}}^{eq}$). Moreover
$Y_{N_i + \bar N_i} \equiv Y_{N_i} + Y_{{\bar N_i}}$ is the total
normalized density of the heavy neutrino $N_i$
 and  $Y_{N_i - \bar N_i} 
\equiv Y_{N_i} - Y_{\bar N_i}$ 
is the corresponding $L_N$ asymmetry.

To write the Eqs.~\eqref{eq:be1-1}--\eqref{eq:be1-3} we have also defined the 
 following combinations of reaction densities: 
\begin{eqnarray}
[a,b,...\leftrightarrow i,j,...]&=& \frac{n_a}{n_a^{eq}}
\frac{n_b}{n_b^{eq}} \gamma^{eq}(a,b,...\rightarrow i,j,...) -
\frac{n_i}{n_i^{eq}} \frac{n_j}{n_j^{eq}}
\gamma^{eq}(i,j,...\rightarrow a,b,...), \\
(a,b,...\leftrightarrow i,j,...) &\equiv& [a,b,...\leftrightarrow
i,j,...] - [\bar{a},\bar{b},...\leftrightarrow \bar{i},\bar{j},...] \; ,\\
\{a,b,...\leftrightarrow i,j,...\} &\equiv&
[a,b,...\leftrightarrow i,j,...] + [\bar{a},\bar{b},...\leftrightarrow
\bar{i},\bar{j},...]  \; ,
\end{eqnarray}
and the prime written in the contribution of some processes indicates that
the on-shell contribution to them has to be subtracted. Note
that we have not included 
scatterings involving gauge bosons. They do not
introduce qualitatively new effects and no further density asymmetries
are associated to them. We have also
ignored finite temperature corrections to the particle masses and
couplings \cite{gi04}. In particular we take all equilibrium number
densities $n_X^{eq}$, with $X \neq N_i, \bar N_i$, equal to those of 
massless particles.

After summing over the most relevant contributions
\footnote{
The scatterings mediated by the Higgs or leptons involving two external
heavy neutrinos have been neglected because they are much slower than
the scatterings
involving only one external heavy neutrino and the top quark, since 
the Yukawa couplings of the heavy neutrinos are much smaller than
the Yukawa of the top quark.}
we find: 
\begin{eqnarray}
\frac{\dif Y_{N_i + \bar N_i}}{\dif z} &=&\frac{-2}{sHz}
\left(\frac{Y_{N_i + \bar N_i}}{Y_{N_i + \bar
N_i}^{eq}}-1\right) \sum_\alpha \left(
\g{N_i}{\ell_\alpha h}+ \g{N_i\bar \ell_\alpha}{\bar Q_3 t}
+ 2 \,
\g{N_i Q_3}{t \ell_\alpha} \right) \; ,  
\label{eq:be2}\\
\frac{\dif Y_{N_i - \bar N_i}}{\dif z} &=& \frac{-1}{sHz} \left\{
\sum_\alpha \g{N_i}{\ell_\alpha h} \left[ y_{N_i} - y_{\ell_\alpha} - y_h \right] + \g{N_i \bar \ell_\alpha}{\bar Q_3 t} \left[ y_{N_i} - \frac{Y_{N_i + \bar N_i}}{Y_{N_i + \bar N_i}^{eq}} y_{\ell_\alpha} + y_{Q_3} - y_t \right]
\right. \nonumber
\\ 
&& \left. + \sum_\alpha 
\g{N_i Q_3}{t \ell_\alpha} \left[ 2 y_{N_i} - 2 y_{\ell_\alpha}
+ \left( 1 + \frac{Y_{N_i + \bar N_i}}{Y_{N_i + \bar N_i}^{eq}} \right) \left(y_{Q_3}-  y_t \right) \right] \right\} , \label{eq:be3} 
\\
\frac{\dif Y_{\Delta_\alpha}}{\dif z} & =& \frac{-1}{sHz} \left\{
\sum_i \left( \frac{Y_{N_i + \bar N_i}}{Y_{N_i + \bar N_i}^{eq}} - 1
\right)\epsilon_{\alpha i} \, 2 \sum_\beta \left(\g{N_i}{\ell_\beta h} 
+ \g{N_i \bar \ell_\beta}{\bar
Q_3 t} + 2 \, \g{N_i Q_3}{t
\ell_\beta} \right) \right. \nonumber \\ 
&& \left.  + \sum_i \g{N_i}{\ell_\alpha h} \left[ y_{N_i} -
y_{\ell_\alpha} - y_h \right] + \g{N_i \bar \ell_\alpha}{\bar Q_3 t}
\left[ y_{N_i} - \frac{Y_{N_i + \bar N_i}}{Y_{N_i + \bar N_i}^{eq}}
y_{\ell_\alpha} + y_{Q_3} - y_t \right] \right.\nonumber \\ 
& & \left. + \sum_i
\g{N_i Q_3}{t \ell_\alpha} \left[ 2 y_{N_i} - 2 y_{\ell_\alpha}
+ \left( 1 + \frac{Y_{N_i + \bar N_i}}{Y_{N_i + \bar N_i}^{eq}} \right) \left(y_{Q_3}-  y_t \right) \right] \right. \nonumber\\ 
&& \left.  + \sum_{\beta \neq
\alpha} \left( 
\gp{\ell_\beta h}{\ell_\alpha h} + 
\g{\ell_\beta \bar{h}}{\ell_\alpha \bar{h}} + 
\g{h\bar{h}}{\ell_\alpha\bar{\ell_\beta}} 
\right) [y_{\ell_\beta} - y_{\ell_\alpha}] \right\} \; ,
\label{eq:be4}
\end{eqnarray} 
where we have introduced the notation 
$\g{a, b, \dots}{c, d, \dots}
\equiv \gamma(\proname{a, b, \dots}{c, d, \dots})$.

In writing Eqs.~\eqref{eq:be3} and \eqref{eq:be4} we have accounted
for the fact  that  the CP asymmetry in scatterings is equal to the CP
asymmetry in decays since only the wave part contributes to the CP
asymmetry of the different processes (see~\cite{nrr}). Moreover, since the 
total CP asymmetries $\epsilon_i$ are 
null, there are no source terms proportional to $\epsilon_{\alpha i}$ in the 
equation for the evolution of $Y_{N_i - \bar N_i}$, which is therefore driven 
only by terms proportional to the different density asymmetries.

It is also important to notice that the equations for the asymmetries are 
not all independent
due to the condition $\frac{\dif Y_{B - L_{SM} - L_N}}{\dif z}=0$, where
$Y_{B - L_{SM} - L_N} \equiv Y_B - Y_{L_{SM}} - Y_{L_N}$, 
$Y_{L_{SM}} \equiv \displaystyle\sum_\alpha Y_{L_\alpha}$, and 
$Y_{L_N} \equiv \displaystyle \sum_i Y_{N_i - \bar N_i}$. If we take as initial
condition that all the asymmetries are null, then $\displaystyle \sum_\alpha
Y_{\Delta_\alpha} - \sum_i Y_{N_i - \bar N_i} = 0$.

In Fig.~\ref{fig:rates} we plot the different reaction densities
included in the Boltzmann equations, normalized to $H n_\ell^{eq}$,
where $H$ is the expansion rate of the universe.
This normalization is appropriate to study the contribution of the
different processes to the LFVW. We show the figure for
$M_1= 250$  GeV, $M_2 = 275$ GeV, 
$(\lambda^\dag \lambda)_{11}=  8.2 \times 10^{-15} \;  
(\tilde m_1=10^{-3}\; {\rm eV})$, and 
$(\lambda^\dag \lambda)_{22}=  10^{-4}$, 
which are the values of the parameters of one of the examples given in the 
next section.
For other values of the Yukawa couplings the reaction densities
$\g{N_i}{\ell_\alpha h}$, 
$\g{N_i\bar \ell_\alpha}{\bar Q_3 t}$, and 
$\g{N_i Q_3}{t \ell_\alpha}$ scale as 
$(\lambda^\dag \lambda)_{ii}$ while 
$\g{\ell_\beta \bar{h}}{\ell_\alpha \bar{h}}$ and  
$\g{h\bar{h}}{\ell_\alpha\bar{\ell_\beta}}$ scale as
$((\lambda^\dag \lambda)_{22})^2$
\footnote{The contribution of the virtual $N_1$ to the 
LFVW processes is 
negligible in all the cases we are going to deal with, 
so we have only included $N_2$ virtual scatterings.}. 
The subtracted $s$-channel reaction density 
$\gp{\ell_\beta h}{\ell_\alpha h}$ is of the same order 
and scales as the $t$- and $u$-channel ones shown in the plot.
The infrared divergence of the reaction mediated by the Higgs 
in the t-channel has been regularized using a Higgs mass equal to 
$M_1$ in the propagator.
For convenience we have 
factorized out the flavour projection factors as explicitly  
displayed in the figure.

\FIGURE[ht]{
\centering
\includegraphics[width=0.8\textwidth]{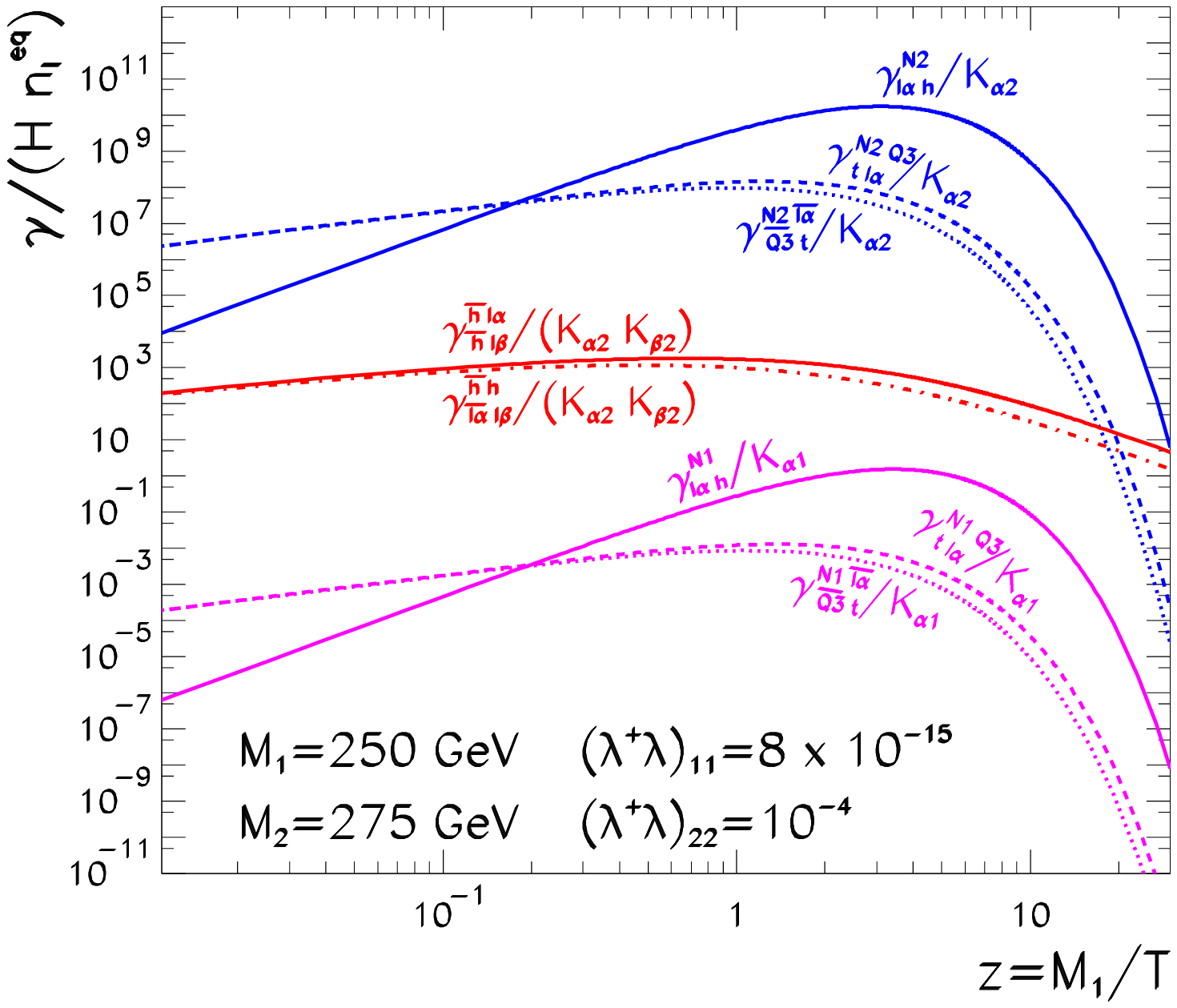}
\caption[]{Reaction densities 
included in the Boltzmann equations as a function of $z=M_1/T$, normalized 
to $H n_\ell^{eq}$, where $H$ is the expansion rate of the universe. 
\label{fig:rates}}}

From the figure we see that, as explained in
Sec.~\ref{sec:requirements}, the rates of processes involving $N_2$
are generically very large (if the CP asymmetry in $N_1$ decays is
required to be not too small). When $M_2 \sim M_1$ the LFVW
induced by processes involving external $N_2$ are dominant over
the ones mediated by virtual $N_2$ in the temperature range relevant
for $N_1$ leptogenesis. Conversely, if there is some hierarchy between
$M_2$ and $M_1$ the LFVW mediated by processes involving virtual
$N_2$ can become the most important ones, since they are not Boltzmann
suppressed for $T > M_2$. 

The network of equations~\eqref{eq:be2}--\eqref{eq:be4} can be
solved after the densities $y_{\ell_\alpha}$, $ \ y_h$
and $y_t-y_{Q_3}$ are expressed in terms of the quantities
$Y_{\Delta_\alpha}$ with
the help of the equilibrium conditions imposed by the fast reactions
which hold in the considered temperature regime 
(see~\cite{spect} and also~\cite{flavour2}). For 
$10^6$ GeV $\gg T\gtrsim T_c$ (where we denote by $T_c$ the critical 
temperature of the electroweak phase transition) they read~\cite{flavour2}
\begin{equation}
y_{\ell_\alpha}= -\sum_\beta C^\ell_{\alpha \beta}\>
\frac{Y_{\Delta_\beta}}{Y^{eq}}, \qquad
\qquad y_h = - \sum_\alpha C^H_\alpha\, 
\frac{Y_{\Delta_\alpha}}{Y^{eq}}\, ,
\label{eq:equil}
\end{equation}
with 
\begin{equation}
C^H = \frac{8}{79}(1 ,\> 1,\> 1) \qquad \hbox{\rm and} \qquad 
C^\ell =\frac{1}{711} \begin{pmatrix}
221 & -16 & -16\\
-16 & 221 & -16 \\
-16 & -16 & 221 \end{pmatrix}\,.
\label{eq:chcl}
\end{equation}
Moreover, the equilibrium condition for the Yukawa interactions of the top quark implies $y_t - y_{Q_3}=y_h/2$.

Above the critical temperature, 
fast sphaleron processes convert the generated lepton asymmetry to
baryon asymmetry \cite{krs}.  Below $T_c$ 
the Higgs starts to acquire its vev and this $SU(2)_L$-breaking
suppresses the sphaleron rate, $\Gamma_{\Delta
(B+L)}$ \cite{clmcw,amc,ls}.  
For temperatures $M_W(T)\ll T \ll M_W(T)/\alpha_W$,  
$\Gamma_{\Delta (B+L)} \sim M_W (M_W(T)/\alpha_W T)^3(M_W(T)/T)^3 \exp
[-E_{sp}/T]$ \cite{amc,resonant}
where $\alpha_W$ is the $SU(2)_L$ fine structure constant, 
$M_W(T)=g v(T)/\sqrt{2}$ is the
$W$-boson mass and the sphaleron energy is $E_{sp}\sim
M_W(T)/\alpha_W$. Because of the exponential suppression of 
$\Gamma_{\Delta (B+L)}$ the lepton asymmetry is not longer converted
into baryon asymmetry below some temperature $T_f$ for which  
$\Gamma_{\Delta(B+L)}(T_{f})/H(T_f)\leq 1$.

In order to  properly account for the evolution of the relevant
abundances in the temperature regime $T_f<T<T_c$ one must extend
the system of Boltzmann Equations to include the temperature
dependent sphaleron rate. The overall effect can be approximated by 
replacing the  usual conversion factor  $n_B=\frac{28}{79}n_{(B-L_{SM})}$
by a temperature-dependent rate given by \cite{ls,ht}
\begin{equation}
Y_B(T)=4\frac{77T^2+54v(T)^2}{869T^2+666v(T)^2} 
\sum_\alpha Y_{\Delta_\alpha}(T),
\end{equation}
where $v(T)$ is the temperature-dependent Higgs vev: 
\begin{equation}
v(T)=v\left(1-\frac{T^2}{T_c}\right)^{\frac{1}{2}}
\;\;\; 
{\rm with}
\;\;\; T_c=v\left(\frac{1}{4}+\frac{{g'}^2}{16 \lambda}
+\frac{3{g}^2}{16 \lambda}+\frac{\lambda_t}{4 \lambda}\right)^{-\frac{1}{2}} \; . 
\end{equation}
Here $\lambda$ is the quartic Higgs self-coupling, 
$g$ and $g'$ are the $SU(2)_L$ and $U(1)_Y$ gauge couplings,
and $\lambda_t$ is the top Yukawa coupling. 

Since the sphaleron processes are effectively switched off at
$T<T_f$, the baryon asymmetry is unaffected below this temperature.
 
In principle an additional effect is that the set of equilibrium
conditions leading to Eq.~\eqref{eq:chcl} are also modified in the
temperature range between $T_c$ and $T_f$. 
We have verified that this effect does not lead to any
relevant change in our conclusions. Furthermore for  $T<T_c$
the effects of the non-vanishing $v(T)$ must be accounted
for in the reaction densities. As long as $M_{i}$ is large
enough compared to  $T_c$ these effects can be safely neglected. 

\section{Results}
\label{sec:results}

In order to quantify the required conditions for generating the
observed baryon asymmetry, 
$Y_B = (8.75 \pm 0.23) \times 10^{-11}$ ~\cite{lastwmap}, 
we have solved the Boltzmann equations presented in the previous section. 
As discussed at the end of Sec.~\ref{sec:iss}, the CP asymmetry in 
processes involving $N_1$ or $N_2$  is larger the  closer the masses 
of $N_2$ and $N_1$  are to each other. Indeed the proximity of the masses 
of the heavy neutrinos is the key parameter that determines whether or 
not successful leptogenesis is possible.
We first show how successful leptogenesis  is possible in
this scenario in the resonant mass regime. We then 
explore the requirements for obtaining the observed baryon asymmetry without 
reaching the resonant condition.

\subsection{Resonant case}
In the case that the CP asymmetry of $N_1$ decays receives a resonant
contribution from $N_2$ the proposed mechanism works in a wide range 
of the parameter space.

As an illustration we show in Fig.~\ref{fig:res} an 
explicit example 
in which  
\begin{equation}
\begin{split}
M_1& =800~{\rm GeV}\;, \quad  
M_2 = M_1 + \frac{\Gamma_{N_2}}{2} \;, \\
(\lambda^\dag \lambda)_{11} &= 10^{-12}\; ,\quad (\lambda^\dag \lambda)_{22}
= 10^{-10}\;,\\
 K_{e 1}&= 0.3, \quad K_{\mu 1}=0.3, \quad K_{\tau 1}=0.4, \\
K_{e 2}&= 0.1, \quad K_{\mu 2}=0.1, \quad K_{\tau 2}= 0.8 \; ,\\ 
p^{12}_{e \mu}&=p_{e \tau}^{12}=p_{\mu \tau}^{12}=1 .
\end{split}
\label{eq:resparam}
\end{equation}
With these values the corresponding CP asymmetries are:
$\epsilon_{e 1} = 6.5 \times 10^{-2},\epsilon_{\mu 1} = 3.5 \times
10^{-2}, \epsilon_{\tau 1} = -1 \times 10^{-1}$, $\epsilon_{e 2} = 1.3
\times 10^{-3},\epsilon_{\mu 2} = 7 \times 10^{-4}$, and
$\epsilon_{\tau 2} = - 2 \times 10^{-3}$.

In deriving the final baryon asymmetry we have taken  
$m_H=200$ GeV for which $T_c\simeq 150$ GeV and $T_f\simeq 100$ GeV. 

\FIGURE[ht]{
\centerline{\protect\hbox{
\epsfig{file=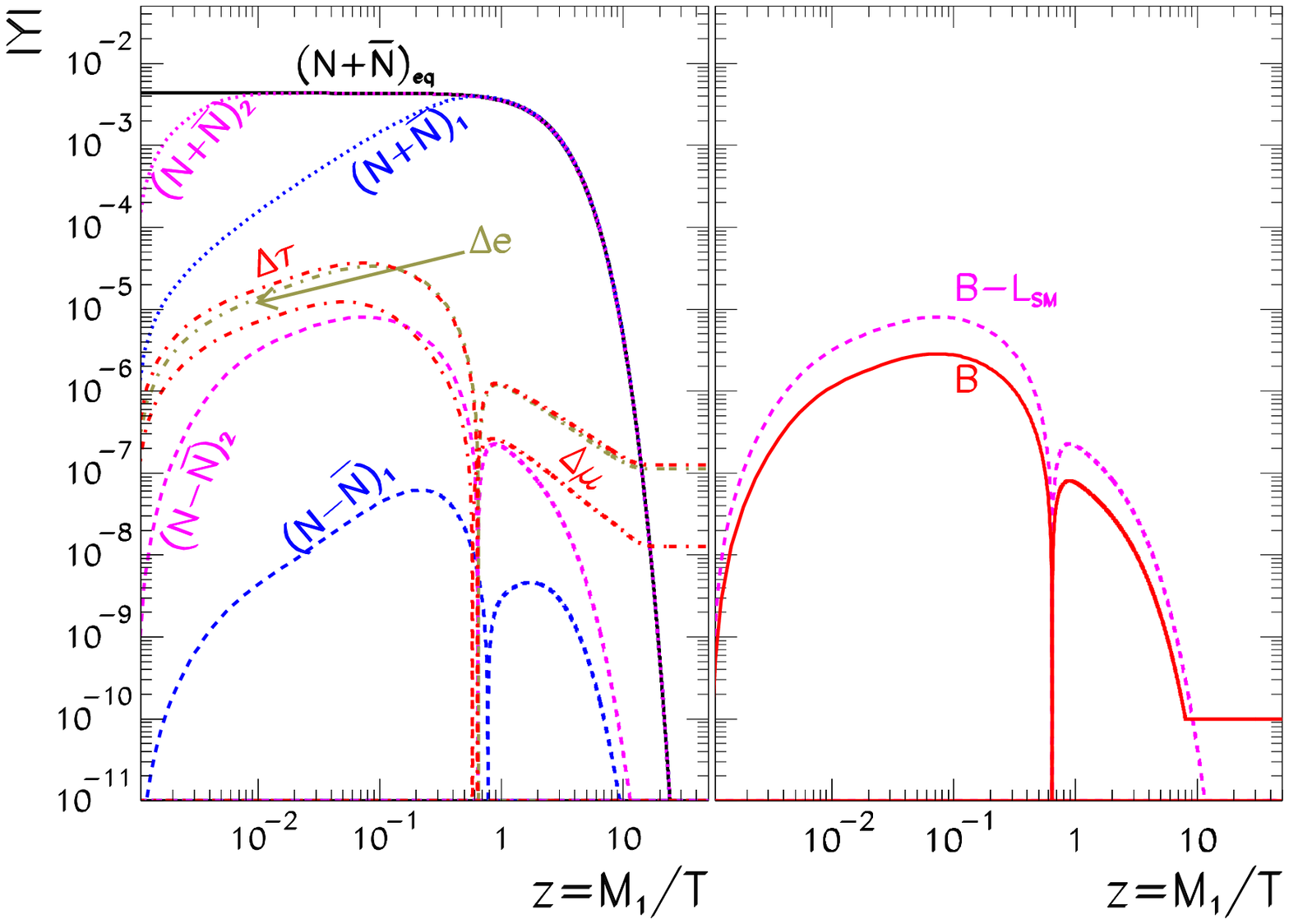,width=\textwidth}}}
\caption[]{The asymmetries $|Y_{\Delta_\alpha}|, |Y_{N_i - \bar N_i}|,
Y_{B-L_{SM}}$, and the densities $Y_{N_i + \bar N_i}, Y_{N_i + \bar
N_i}^{eq}$ as a function of $z$ 
for the values of
the parameters given in Eq.~\eqref{eq:resparam}. 
\label{fig:res}}}

The figure explicitly shows that  $Y_{B-L_{SM}}\rightarrow 0 $ as 
$T\rightarrow 0$   which is mandatory due 
to the conservation of $B-L$ in this scenario and the assumed 
initial conditions.    
Indeed it can be verified that, although $Y_{\Delta_\alpha}$
saturate at a finite value at low temperature, 
with  the sign assignments for the asymmetries, at any $z$ it is verified that
$Y_{B-L_{SM}}=|Y_{\Delta_e}|
+|Y_{\Delta_\mu}|-|Y_{\Delta_\tau}|= Y_{L_N} =
|Y_{N_1-\overline{N_1}}|+|Y_{N_2-\overline{N_2}}|
\simeq |Y_{N_2-\overline{N_2}}|$. 
Still as illustrated in the figure, 
the observed baryon asymmetry is generated 
once the sphalerons  switch off below $T_f=100$ GeV ($z>8$). 

 We notice that despite the CP asymmetry $\epsilon_{\alpha 1}\gg
\epsilon_{\alpha 2}$, the $N_i$ asymmetries verify
$|Y_{N_1-\overline{N_1}}|\ll|Y_{N_2-\overline{N_2}}|$.  This is so
because, even though the lepton asymmetries $y_{\ell_\alpha}$ are mostly
produced in processes involving $N_1$, it is the inverse decay $\ell_\alpha
h\rightarrow N_i$ what determines how much of the lepton asymmetries
is transferred to the $N_i$ asymmetry. 
Moreover, recall that there are no source
terms proportional to $\epsilon_{\alpha i}$ in the evolution equations
of $Y_{N_i-\overline{N_i}}$ Eq.~\eqref{eq:be3}. Then,   
since the $N_2$ Yukawa
couplings are larger, the inverse decays of the $N_2$ are more
efficient and a larger $N_2$ asymmetry is produced. Consequently 
for the small $N_1$ Yukawa couplings considered, the $N_1$ asymmetry plays a
very little role in the dynamics of the system.

We note also that contrary to leptogenesis scenarios which occur well above 
the electroweak scale, once the heavy neutrinos have decayed, the universe is 
left with an equal amount of lepton and baryon asymmetry. 
Furthermore the flavour 
asymmetries $Y_{\Delta \alpha}$ typically remain some orders of magnitude 
greater than $Y_B$.

\subsection{Non-resonant case}
\FIGURE[ht]{
\centerline{\protect\hbox{
\epsfig{file=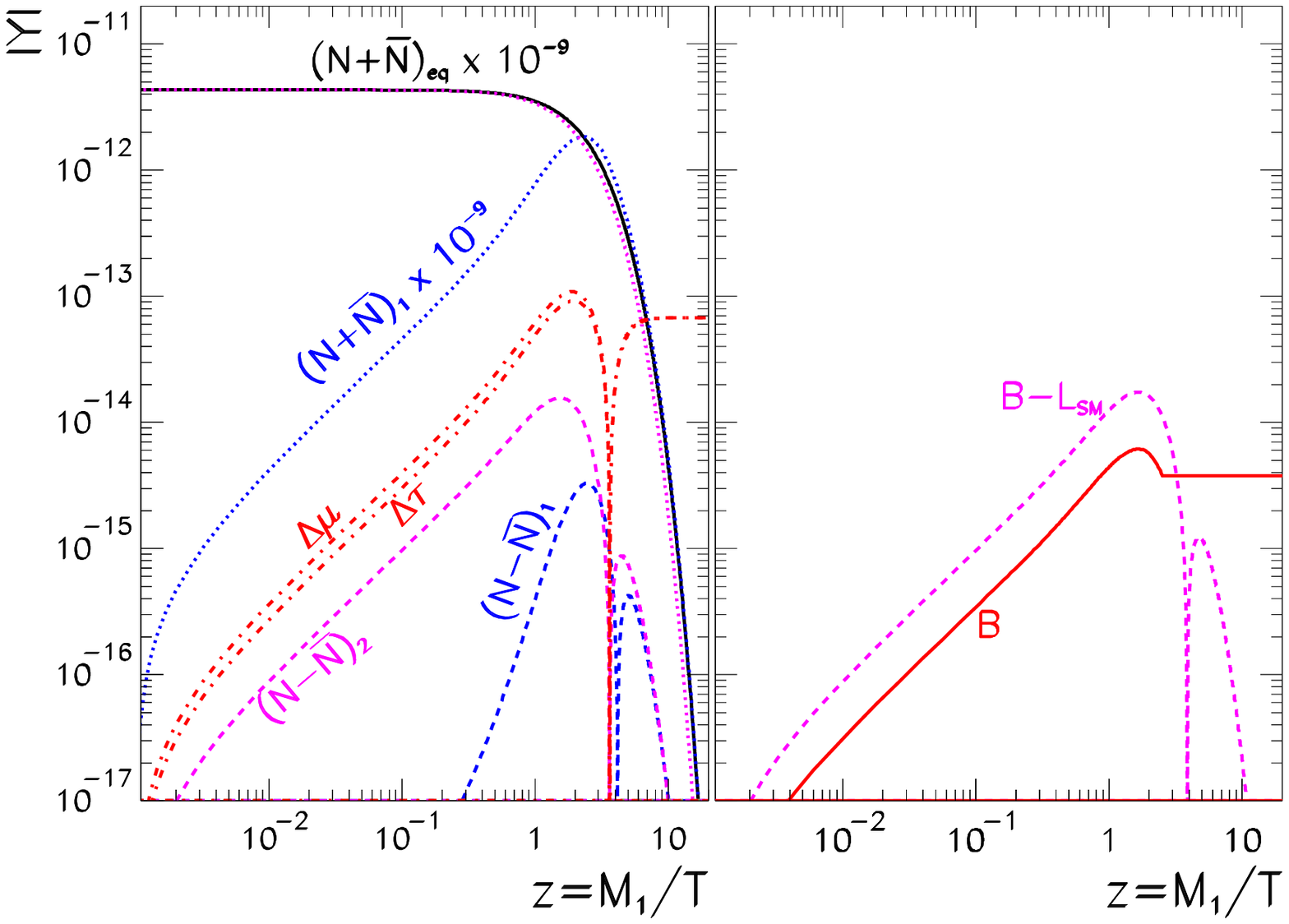,width=\textwidth}}}
\caption[]{The asymmetries $|Y_{\Delta_\alpha}|, |Y_{N_i - \bar N_i}|,
Y_{B-L_{SM}}$, and the densities $Y_{N_i + \bar N_i}, Y_{N_i + \bar
N_i}^{eq}$ as a function of $z$ for the values of
the parameters given in Eq.~\eqref{eq:noresparam}. 
Note that the densities $Y_{N_i + \bar N_i}$ and 
$Y_{N_i + \bar
N_i}^{eq}$  have been rescaled by $10^{-9}$ 
in order to fit into the figure.  
\label{fig:nores}}}

In this regime, in order to have enough CP asymmetry 
at least one of the Yukawas of $N_2$ has to be very large. 
The strongest bounds on the Yukawas of the heavy neutrinos for the range 
of masses we are interested in come from
constraints on  violation of weak universality, lepton flavour violating 
processes and collider signatures \cite{unitlim} which allow for 
$|(\lambda^\dag \lambda)_{22}| \lesssim 5 \times 10^{-3} (M_2/v)^2$. 

However for flavour effects to be  relevant, the Yukawas of the $N_2$ must 
be smaller than the Yukawas of (at least) one of the charged 
leptons, so we require that 
$\lambda_{2\alpha}\lesssim \lambda_\tau \sim 10^{-2}$~\footnote{This 
is a rough estimate of the validity of the "fully flavor 
regime" which is
enough for our purposes; for a more detailed analysis about this point
see~\cite{zeno}.}.
This is so because if the fastest leptonic reaction  rates were those 
associated to  $N_2$, the flavour basis which diagonalizes the density 
matrix would be that formed by the lepton in which $N_2$ decays, 
$\ell_2=\sum_\alpha  \lambda_{\alpha 2} \ell_\alpha
/\sum_\alpha| \lambda_{\alpha 2} |^2$,
together with two states orthogonal to $\ell_2$ and hence no baryon 
asymmetry would be generated since in this basis the CP 
asymmetry is zero.

Furthermore if the Yukawa couplings of
$N_2$ with all the three light leptons were comparable
and large (even if smaller than $\lambda_\tau$) there would be strong
LFVW in all flavours (see Fig.~\ref{fig:rates}) 
and the generation of the baryon asymmetry would be strongly suppressed.  
Therefore, as was explained before, some of the Yukawa couplings have to 
be very small. Note that for a fixed value of $(\lambda^\dag \lambda)_{22}$ 
the CP asymmetry in a given flavour decreases linearly
with decreasing Yukawa coupling of that flavour with $N_2$, 
while the washout terms decrease quadratically with it. We conclude that the
most favorable situation is to have a strong hierarchy in the flavour
structure of the $N_2$ Yukawas.
 
Concerning the optimum range of Yukawa couplings for $N_1$ there are
two relevant effects. On one hand if they are too large -- 
$\tilde m_1 \gtrsim m_* \simeq 10^{-3}$~eV -- the maximum values of the 
$B-L_{SM}$ asymmetry occur at $z < 1$, so
that $M_1<T_f$ in order for the sphalerons to decouple when the baryon
asymmetry is largest. In this case the analysis is more complex because one 
cannot neglect the effects of the breaking of $SU(2)$ in the reaction 
densities. The expected effect is the reduction of the $N_1$ decay rate due 
to the phase space factors. In order to determine the effect on the final 
$B$ asymmetry a dedicated study is required which is beyond the scope of 
this paper. On the other hand, for values
$\tilde m_1\ll m_*$, the peak in the $B-L_{SM}$
asymmetry shifts to large values of $z$ but it is in general lower
because of the smaller production of $N_1$ when starting from a zero 
abundance as initial condition.  One may wonder if this
conclusion may be modified when assuming a non-vanishing initial $N_1$
abundance which would allow for very late $N_1$ decay. It is not,
because at large values of $z$ the LFVW are very suppressed and therefore 
the flavour effects, which are essential in this scenario, do not survive.
In summary, generically larger baryon asymmetries are expected for 
$\tilde m_1\sim m_*$.

 With the above considerations in mind, we have explored the parameter space 
for a fixed value of $M_2/M_1$ (chosen near to 1) and in 
Fig.~\ref{fig:nores} we plot the
evolution of the different asymmetries and densities for a set of parameters 
representative of the cases with highest production of baryon asymmetry:
\begin{equation}
\begin{split}
M_1 &= 250\; {\rm GeV},\quad M_2 = 275 \; {\rm GeV}     \; , \\
(\lambda^\dag \lambda)_{11} &=  8.2 \times 10^{-15} \;  
(\tilde m_1=10^{-3}\; {\rm eV})\; , \\
(\lambda^\dag \lambda)_{22} &=  10^{-4} \; ,\\
K_{e 1} &= 0.\; , \quad K_{\mu 1}=0.3\;, \quad K_{\tau 1}=0.7\;, \\
K_{e 2} &= 0.\;, \quad K_{\mu 2}=10^{-10}\;, \quad K_{\tau 2}\simeq 1 \; ,\\ 
p^{12}_{e \mu} &= p_{e \tau}^{12}=p_{\mu \tau}^{12}=1\; .
\end{split}
\label{eq:noresparam}
\end{equation}
Since $\sqrt{(\lambda^\dag \lambda)_{22}} > \lambda_\mu$ 
we have safely chosen the projectors $K_{e 1}, K_{e 2}=0$ to prevent
any possible flavour projection effects associated to the Yukawa interactions 
of $N_2$, which would complicate the description of the problem without 
substantially changing the results. In an effectively two flavour case the CP 
asymmetries are proportional to only one phase factor, in this case to 
$p_{\mu \tau}^{12}$, therefore we have adopted in the example its 
maximum possible value.
With these values the corresponding CP asymmetries are:
$\epsilon_{e 1} = \epsilon_{e 2} =0 ,
\epsilon_{\mu 1} = -\epsilon_{\tau 1} = 8.7 \times 10^{-11}, 
\epsilon_{\mu 2} = -\epsilon_{\tau 2} = 8.7 \times 10^{-21}$.


From the figure it can be seen that even with these large 
$N_2$ Yukawa couplings, 
their strong flavour hierarchy  and the small $1/(a_2-1)$ suppression, 
the produced $Y_B$ falls short to explain the observations 
by about 5 orders of magnitude.
However, we notice that in this regime the asymmetry comes mainly from 
processes involving $N_1$. Therefore the $B-L_{SM}$ asymmetry is approximately 
proportional 
to $1/(a_2-1)$ which is around 5 in the example. If we take $M_2$ closer
to $M_1$, that factor and the corresponding $Y_B$   
grow accordingly. Thus we see that in this regime it is also
possible to generate the required baryon asymmetry as long
as $N_1$ and $N_2$ are strongly degenerated even if still not in
the resonant regime. For example for the values
of parameters given in Eq.~\eqref{eq:noresparam}  the observed
baryon asymmetry could be produced if  $a_2-1 \sim 2.4\times 10^{-5}$,  
which is still far from the resonance ($g_2 = 4 \times 10^{-6}$).

Something to note is that despite the great hierarchy among the
projectors of $N_2$ onto $\ell_\mu$ and $\ell_\tau$, the flavour
asymmetries $Y_{\Delta \alpha}$ ($\alpha = \mu, \tau$) are quite
similar in size. This is because the evolutions of the different
asymmetries are strongly coupled due to the conservation of $B-L$ and
the null value of the total CP asymmetry in $N_1$ decays.

To obtain an estimate of the order of magnitude of the density asymmetries as well as to understand their dependence on the $N_2$  
flavour projectors (which are more relevant than the $N_1$ projectors when 
$(\lambda^\dag \lambda)_{22} \gg (\lambda^\dag \lambda)_{11}$), we have 
developed the following semiquantitative approximation: 

(i) From Fig.~\ref{fig:rates} we  see that for $K_{\mu 2} \gtrsim 10^{-9}$ 
the rates of the processes $\lrproname{N_2}{\ell_{\mu, \tau} h}$ are larger 
than the expansion rate of the Universe in the most relevant range of 
temperatures. Therefore at each instant the thermal bath has time to relax, 
i.e., the production of asymmetry equals its erasure. 
This implies that the derivatives of the density asymmetries are negligible 
with respect to the source and LFVW terms, hence we
set them to zero in the Boltzmann equations. 

(ii) We keep the CP asymmetries produced by $N_1$ 
(since $\epsilon_{\alpha 2} \ll \epsilon_{\alpha 1}$), 
as well as the $N_2$ decay and inverse decay reactions, and
neglect all the remaining subdominant contributions, including 
the partial conversion of lepton asymmetry into baryon asymmetry during the 
leptogenesis era.


Within this approximation,  Eqs.~\eqref{eq:be3} and \eqref{eq:be4} simplify to
\begin{eqnarray}
S_\mu(z) + K_{\mu 2} \g{N_2}{\ell h} \left[ y_{N_2} - y_{\ell_\mu} \right] &=& 0\; , \\
S_\tau(z) + K_{\tau 2} \g{N_2}{\ell h} \left[ y_{N_2} - y_{\ell_\tau} \right] &=& 0 \; ,
\end{eqnarray}
where $S_\mu(z) = -S_\tau(z) =  \epsilon_{\mu 1} \, \bigl(\tfrac{Y_{N_1 + \bar N_1}}{Y_{N_1 + \bar N_1}^{eq}}-1 \bigr) \, 2 \, \g{N_1}{\ell h}$ is the source term normalized to $(-sHz)^{-1}$ and $ \g{N_i}{\ell h} \equiv \sum_\beta  \g{N_i}{\ell_\beta h} $. 

A third equation is provided by total lepton number conservation,  
i.e. $Y_{N_2 - \bar N_2} + Y_{L_\mu} + Y_{L_\tau} = 0$. In the most relevant temperature 
range for the model, $T \sim M_2$, we can approximate the equilibrium density of $N_2$ by 
that of one relativistic degree of freedom, therefore 
\begin{equation}
y_{N_2} + 2y_{\ell_\mu} + 2y_{\ell_\tau} = 0 \; .
\end{equation}

When $K_{\mu 2} \ll 1 \simeq K_{\tau 2}$ the solution to this system of 
equations is 
\begin{eqnarray}
y_{\ell_\mu}  &=& \frac{3}{5} \frac{S_{\mu}(z)}{K_{\mu 2} \g{N_2}{\ell h}}\; , 
\nonumber \\
y_{N_2}  = y_{\ell_\tau} &=& 
- \frac{2}{5} \frac{S_{\mu}(z)}{K_{\mu 2} \g{N_2}{\ell h}}\; .
\end{eqnarray}


From this analysis, 
it is clear that despite the large hierarchy in the projectors of $N_2$, all the density asymmetries have the same order of magnitude. Moreover, since the source term is proportional to $\sqrt{K_{\mu 2}}$, the density asymmetries are inversely proportional to $\sqrt{K_{\mu 2}}$. We have verified numerically that this dependency of the asymmetries on the projector actually holds in the range $10^{-9} \lesssim K_{\mu 2} \lesssim 10^{-1}$, where the approximations we made are expected to be valid. 

For $\sqrt{K_{\mu 2}} \lesssim 10^{-10}$ the rates of the processes $\lrproname{N_2}{\ell_{\mu} h}$ are lower than the expansion rate of the Universe, hence point (i) is not longer true. In this range of $K_{\mu 2}$, the density asymmetries decrease as $\sqrt{K_{\mu 2}}$ because the main dependence on $K_{\mu 2}$ comes from the relation $\epsilon_{\mu,\tau \, 1} \propto \sqrt{K_{\mu 2}}$. Thus, 
fixing all the parameters but $K_{\mu 2}$ to the values given in Eq.~\eqref{eq:noresparam}, the baryon asymmetry is maximized for $K_{\mu 2}$ between $10^{-9}$ and $10^{-10}$.

\section{Summary}
In this article we have studied the possibility of generating the observed
baryon asymmetry via leptogenesis in the decay (and scatterings)
of heavy Dirac Standard Model singlets  with ${\cal O}$ (TeV) masses
in a framework with $B-L$ conservation above the electroweak scale.
In this scenario a total lepton number,  which is perturbatively 
conserved, can be  defined. This lepton number  is shared between the 
Standard Model leptons and the heavy Dirac singlets as described in
Sec.\ref{sec:iss}. In this scenario, despite the total CP
asymmetry is null, a CP asymmetry in the different
SM lepton flavours can be generated (see Sec.\ref{sec:CP}).

The additional physical condition for generating a non-vanishing $B$ in 
this framework is that the sphalerons depart from equilibrium during the
decay epoch. For symmetric initial conditions  (no net baryon nor total lepton 
number present), $B-L_{SM} = 0$ after the heavy neutrinos have 
disappeared. Consequently if the sphalerons were still active after the
heavy Dirac singlets decay epoch, the final baryon asymmetry, being
proportional to $B-L_{SM}$, would be zero. However if they depart from
equilibrium
during the decay epoch the baryon asymmetry freezes at a value which
in general is not null.

In summary in this scenario the baryon asymmetry is generated by the 
interplay of lepton flavour effects and the sphaleron decoupling in the 
decay epoch.
In order to quantify whether enough baryon asymmetry can be generated we 
have constructed and solved the network of relevant Boltzmann Equations 
associated with the abundances of the two lightest Dirac heavy singlets, 
their asymmetries and the three SM flavour asymmetries. The results are 
given in Sec.\ref{sec:results}.

We find that the ratio of the masses of 
the two heavy Dirac neutrinos is the key parameter that determines whether 
or not successful leptogenesis is possible. The relevant Yukawa couplings 
are constrained from above by the requirement of
having flavour effects, and from 
below by the requirement of large enough CP asymmetry. Within these 
boundaries we conclude that successful leptogenesis can occur if the 
masses of two heavy Dirac singlets are very degenerate $M_2/M_1-1 \lesssim 
{\cal O}(10^{-5})$. 
Recall that in our framework this degeneration is not a consequence of 
the small breaking of total lepton number, as in other low scale leptogenesis 
scenarios \cite{resonant,ab}, although it may be due to an additional 
symmetry of the heavy sector.
In particular, if the CP asymmetry is resonantly 
enhanced -- $(M_2^2-M_1^2) \sim M_1 \Gamma_2$ -- , the mechanism works 
for a wide range of values of the Yukawa couplings and flavour 
projections.
It is worth to explore whether the heavy neutrinos will be observable 
at the LHC and/or lead to measurable lepton flavour violating signals,
within the parameter regions allowed by leptogenesis.

\section*{Acknowledgments}
We thank Pilar Hern\'andez for very useful discussions about the
equilibrium conditions for baryogenesis.
This work is supported by spanish  MICCIN  under grants 2007-66665-C02-01, 
FPA-2007-60323, 
and Consolider-Ingenio 2010  CUP (CSD2008-00037), CPAN
(CSD2007-00042) and PAU (CSD2007-00060),
by CUR Generalitat de Catalunya  grant 2009SGR502,
by Generalitat Valenciana grant PROMETEO/2009/116  
and by USA-NSF grant PHY-0653342.

\end{document}